\documentclass[12pt]{article}
\usepackage{epsfig}
\title{ \bf The Mass and Leptonic Decay Constant of $D_{s0}(2317)$ Meson in the framework
of thermal QCD sum rules}

\author{El\c{s}en Veli Veliev *, G\"{u}l\c{s}ah Kaya **
\\Physics Department, Kocaeli University, Umuttepe Yerle\c{s}kesi \\
41380 Izmit, Turkey \\** e-mail: elsen@kocaeli.edu.tr
\\*** e-mail: gulsahbozkir@kocaeli.edu.tr}
\date{}
\begin{document}
\setlength{\baselineskip}{24pt}
\maketitle
\setlength{\baselineskip}{7mm}
\begin{abstract}
In the present work, we assume $D_{s0}(2317)$ meson as the
$c\overline{s}$ state and study its parameters at finite temperature
using QCD sum rules. It is calculated the annihilation and
scattering parts of spectral function in the lowest order of
perturbation theory. Taking into account perturbative two-loop order
$\alpha_{s}$ corrections and nonperturbative corrections up to the
dimension six condensates it is investigated the temperature
dependences of mass and leptonic decay constant of $D_{s0}(2317)$
meson.
\end{abstract}

\setcounter{page}{1}
\section{Introduction}
In 2003 BaBar Collaboration discovered a positive-parity scalar
charm strange meson $D_{s0}(2317)$ with a very narrow width
\cite{1}, which was confirmed by CLEO Collaboration \cite{2} and
BELLE Collaboration \cite{3} later. This observed state has
attracted much attention because its measured mass and width do not
match the predictions from potential-based quark models \cite{4}. To
resolve the difficulties, many theoretical models have appeared in
the literature. Various theoretical models, based on the
$c\overline{s}$ quark structure, are suggested to explain the low
mass and the narrow width for the $D_{s0}(2317)$ meson
\cite{5}-\cite{10}. QCD sum rule analysis in \cite{11}, \cite{12}
supports the $c\overline{s}$ postulation of nature $D_{s0}(2317)$.
Apart from the quark-antiquark interpretation, this state has been
interpreted as a $DK$ molecule \cite{13}, a $D_{s}\pi$ molecule
\cite{14}, a $c\overline{s}q\overline{q}$ four-quark state
\cite{15}, and a mixing of the conventional state and the four-quark
state \cite{16}. Also this state was investigated in the framework
of chiral symmetry considerations \cite{17}.

In this work, we assume $D_{s0}(2317)$ meson as the $c\overline{s}$
state and study its parameters at finite temperature using QCD sum
rules \cite{18}. The extending of QCD sum rules method to finite
temperature has been made in the paper \cite{19}. This extension
based on two basic assumptions, that the Operator Product Expansion
(OPE) and notion of quark-hadron duality remain valid at finite
temperature, but the vacuum condensates must be replaced by their
thermal expectation values. The thermal QCD sum rule has been
extensively used for studying thermal properties of both light and
heavy mesons as a reliable and well-established method
\cite{20}-\cite{24}.

In the present work, we calculated the temperature behavior of mass
and leptonic decay constant of $D_{s0}(2317)$ meson. The knowledge
of leptonic decay constants is needed to predict numerous heavy
flavor electroweak transitions and to determine Standard Model
parameters from the experimental data. Also leptonic decay constants
play essential role in the analysis of CKM matrix, CP violation and
the mixings $\overline{B_{d}}B_{d}$, $\overline{B_{s}}B_{s}$.

This paper is organized as follows. In section 2, we calculated the
annihilation and scattering parts of spectral density and give the
expression for the perturbative scalar spectral function up to
two-loop order $\alpha_{s}$ corrections. Also nonperturbative
contributions up to the dimension six condensates \cite{16} are
summarized. Section 3 contains our numerical analysis of the mass
and leptonic decay constant using Borel transform sum rules.

\section{Thermal QCD sum rule for the scalar charm strange meson}
The starting point for the sum rule analysis is the two-point
thermal correlator
\begin{equation}\label{eqn1}
\Pi(q^{2})=i \int d^{4}x e^{iq\cdot x} \langle T(J(x)J^{+}(0))\rangle, \\
\end{equation}
where $J(x)=(m_{c}-m_{s}):\bar{s}(x)c(x):$ is heavy-light quark
current and has the quantum numbers of the $D_{s0}(2317)$ meson,
$m_{c}$ and $m_{s}$ are charm and strange quark masses respectively.
$s$ quark mass is not neglected throughout this work. Thermal
average of any operator \emph{O} is determined by following way
\begin{equation}\label{eqn2}
\langle O\rangle=Tr e^{-\beta H}O/Tr e^{-\beta H}, \\
\end{equation}
where $H$ is the QCD Hamiltonian, and $\beta=1/T$ stands for the
inverse of the temperature $T$ and traces carry out over any
complete set of states. According to the basic idea of the QCD sum
rule, we must calculate this correlator in terms of the physical
particles (hadrons) and in quark-gluon language, and then equate
both representations. First let us calculate theoretical part of the
correlator Eq. (1). Up to a subtraction polynomial, which depends on
the large $q^{2}$ behavior, $\Pi(q^{2})$ satisfies following
dispersion relation \cite{19}, \cite{21}, \cite{22}
\begin{equation}\label{eqn3}
\Pi(q^{2})=\int ds \frac{\rho (s)} {s+Q^{2}}+ subtractions , \\
\end{equation}
where $\rho (q)=\frac{1}{\pi}Im \Pi(q)\tanh (\frac{\beta q_{0}}{2})$
is spectral density. In order to calculate two-point thermal
correlator in the lowest order of perturbation theory we use quark
propagator at finite temperature \cite{25}
\begin{equation}\label{eqn4}
S_{11}(q)=(\gamma^{\mu}q_{\mu}+m)\Big{(}\frac{1}{q^{2}-m^{2}+i\varepsilon}+2\pi
i n (\omega_{q})\delta(q^{2}-m^{2})\Big{)}, \\
\end{equation}
Here $n(\omega_{q})$  is the Fermi distribution function,
$n(\omega_{q})=[\exp(\beta \omega_{q})+1]^{-1}$ and
$\omega_{q}=\sqrt{\mathbf{q}^{2}+m^{2}}$ . After some calculations
we find that perturbative part of spectral density is given by
\begin{eqnarray}\label{eqn5}
&&\rho_{pert}(q, T)=\int
\frac{d\mathbf{k}}{(2\pi)^{3}}\frac{\omega_{1}^{2}-\mathbf{k}^{2}+\mathbf{k}\cdot\mathbf{q}-\omega_{1}q_{0}+m_{c}m_{s}}{\omega_{1}\omega_{2}}
\nonumber\\&&\times
[(1-n_{1}-n_{2})\delta(q_{0}-\omega_{1}-\omega_{2})+(n_{1}-n_{2})\delta
(q_{0}-\omega_{1}+\omega_{2})].
\end{eqnarray}
Here $\omega_{1}=\sqrt{\mathbf{q}^{2}+m_{c}^{2}}$ and
$\omega_{2}=\sqrt{(\mathbf{k}-\mathbf{q})^{2}+m_{s}^{2}}$ . Note
that spectral density involves two pieces, one is called the
annihilation term, $\rho_{a,pert}(s,T)$, which survives at $T=0$.
Other term is called scattering term, $\rho_{s,pert}(s,T)$, which
vanishes at $T=0$. As can be seen, delta function
$\delta(q_{0}-\omega_{1}-\omega_{2})$ in Eq. (5) gives the first
branch cut, $q^{2}\geq(m_{c}+m_{s})^{2}$ , which coincides with zero
temperature cut and describes the standard threshold for particle
decays. On the other hand, delta function
$\delta(q_{0}-\omega_{1}+\omega_{2})$ in Eq.(5) shows that an
additional branch cut arises at finite temperature,
$q^{2}\leq(m_{c}-m_{s})^{2}$, and this new branch cut corresponds to
particle absorption from the medium. Therefore, delta functions
$\delta(q_{0}-\omega_{1}-\omega_{2})$ and
$\delta(q_{0}-\omega_{1}+\omega_{2})$ in Eq.(5) contribute in
regions $(m_{c}+m_{s})^{2}+\mathbf{q}^{2}\leq q_{0}^{2}\leq\infty$
and $0\leq q_{0}^{2}\leq \mathbf{q}^{2}+(m_{c}-m_{s})^{2}$
respectively. Taking into account these contributions the
annihilation and scattering parts of spectral density in the case
$\mathbf{q}=0$ can be written as
\begin{equation}\label{eqn6}
\rho_{a,pert}(s,T)=\rho_{0}(s)\Big{[}1-n\Big{(}\frac{\sqrt{s}}{2}\Big{(}1+\frac{m_{c}^{2}-m_{s}^{2}}{s}\Big{)}\Big{)}-n\Big{(}\frac{\sqrt{s}}{2}\Big{(}1-
\frac{m_{c}^{2}-m_{s}^{2}}{s}\Big{)}\Big{)}\Big{]} ,\\
\end{equation}
\begin{equation}\label{eqn7}
\rho_{s,pert}(s,T)=\rho_{0}(s)\Big{[}n\Big{(}\frac{\sqrt{s}}{2}\Big{(}1+\frac{m_{c}^{2}-m_{s}^{2}}{s}\Big{)}\Big{)}-n\Big{(}-\frac{\sqrt{s}}{2}\Big{(}1-
\frac{m_{c}^{2}-m_{s}^{2}}{s}\Big{)}\Big{)}\Big{]} ,\\
\end{equation}
Here $\rho_{0}(s)$ is the correlation function in the lowest order
of perturbation theory at zero temperature and given by
\begin{equation}\label{eqn8}
\rho_{0}(s)=\frac{3(m_{c}-m_{s})^{2}}{8\pi^{2}s}q^{2}(s)v^{3}(s) ,\\
\end{equation}
where $q(s)=s-(m_{c}-m_{s})^{2}$ and $v(s)=(1-4m_{s}m_{c}/q(s)
)^{1/2}$ .
The contribution of perturbative two-loop order $\alpha_{s}$
corrections to the spectral density in perturbation theory at zero
temperature can be written as \cite{26}:
\begin{equation}\label{eqn9}
\rho_{1}(s)=\frac{4\alpha_{s}}{3\pi}\rho_{0}(s)f(x) ,\\
\end{equation}
where $x=m_{c}^{2}/s$, $\alpha_{s}=\alpha_{s}(m_{c}^{2})$ and
\begin{equation}\label{eqn10}
f(x)=\frac{9}{4}+2Li_{2}(x)+\ln x \ln(1-x)-\frac{3}{2}\ln(1/x-1)-\ln (1-x)+x \ln(1/x-1)-\frac{x}{1-x}\ln x .\\
\end{equation}
Here $Li_{2}(x)=-\int^{x}_{0} dt\frac{\ln(1-t)}{t}$ is dilogarithm
function. Note that in $\alpha_{s}$ corrections terms the strange
quark mass is set zero, though in numerical analysis, the mass of
the strange quark is taken account. The subtraction terms in Eq. (3)
are removed by using the Borel transformation, therefore we will
omit these terms. The non-perturbative contribution at zero
temperature to the correlator has following form
\begin{eqnarray}\label{eqn11}
&&\Pi_{np}(q^{2})=m_{c}\lambda\langle0|\bar{s}s|0\rangle\Big{[}1-\frac{1}{2}\varepsilon(3-\lambda)-\lambda\varepsilon^{2}(1-\lambda)+\frac{1}{2}\varepsilon^{3}(1+\lambda-4\lambda^{2}+2\lambda^{3})\Big{]}
\nonumber\\&&+\frac{1} {12\pi}
\lambda\langle0|\alpha_{s}G^{2}|0\rangle\Big{[}1-3\varepsilon\Big{(}1-\frac{8}{3}\lambda+2\lambda^{2}-2\lambda(1-\lambda)\ln(\varepsilon\lambda)\Big{)}\Big{]}
\nonumber\\&&+\frac{M_{0}^{2}}{2m_{c}}\langle0|\bar{s}s|0\rangle\lambda^{2}(1-\lambda)
(1-\varepsilon(2-\lambda))-\frac{8}{27}\frac{\pi\rho}{m_{c}^{2}}\alpha_{s}\langle0|\bar{s}s|0\rangle^{2}\lambda^{2}(2-\lambda-\lambda^{2}),
\end{eqnarray}
which arises in the framework of the OPE and parameterized by vacuum
expectation values of quark and gluon fields in the QCD Lagrangian.
In Eq. (11)  $\lambda=m_{c}^{2}/(Q^{2}+m_{c}^{2})$,
$\varepsilon=m_{s}/m_{c}$  and terms are organized according to
their dimension. It is assumed, that the expansion (11) also remains
valid, but the vacuum condensates must be replaced by their thermal
expectation values \cite{19}. For the light quark condensate at
finite temperature we use the results of \cite{27},\cite{28}
obtained in chiral perturbation theory and temperature dependence of
quark condensate in a good approximation can be written as
\begin{equation}\label{eqn12}
\langle\bar{q}q\rangle=\langle0|\overline{q}q|0\rangle\Big{[}1-0.4\Big{(}\frac{T}{T_{c}}\Big{)}^{4}-0.6\Big{(}\frac{T}{T_{c}}\Big{)}^{8}\Big{]} ,\\
\end{equation}
where $T_{c}$ is critical temperature. The low temperature expansion
of a gluon condensate is proportional to the trace of the energy
momentum tensor \cite{29} and can be approximated \cite{23} as
\begin{equation}\label{eqn13}
\langle\alpha_{s}G^{2}\rangle=\langle0|\alpha_{s}G^{2}|0\rangle\Big{[}1-\Big{(}\frac{T}{T_{c}}\Big{)}^{8}\Big{]} .\\
\end{equation}
Also, we have used for the mixed condensate the parameterization
\begin{equation}\label{eqn14}
g\langle\bar{q}\sigma_{\mu\nu}\frac{\lambda_{a}}{2}G_{a}^{\mu\nu}q\rangle=M_{0}^{2}\langle\bar{q}q\rangle \\
\end{equation}
and deduced the value of the QCD scale $\Lambda$ from the value of
$\alpha_{s}(M_{Z})=0.1176$ .

Our next task is the calculation of the physical part of the
correlator (1). According to the basic idea of quark-hadron duality
assumption, the right-hand side of eq. (1) can be evaluated in a
hadron-based picture. Equating OPE and hadron representations of
correlation function and using quark-hadron duality the central
equation of our sum-rule analysis takes the form:
\begin{equation}\label{eqn15}
\frac{f^{2}(T)m^{4}(T)}{Q^{2}+m^{2}(T)}=\int^{s_{0}(T)}_{(m_{c}+m_{s})^{2}} ds\frac{\rho_{a,pert}(s,T)+\rho_{1}(s)}{s+Q^{2}}+\int^{(m_{c}-m_{s})^{2}}_{0} ds\frac{\rho_{s,pert}(s,T)}{s+Q^{2}}+\Pi_{np}(Q^{2},T)  , \\
\end{equation}
where $f$ and $m$ are the leptonic decay constant and mass of
$D_{s0}(2317)$ meson respectively. Note that in Eq.(15) the mass and
leptonic decay constant were replaced by their temperature dependent
values.  The continuum threshold  also depends on temperature; to a
very good approximation its scales universally as the quark
condensate \cite{23}
\begin{equation}\label{eqn16}
s_{0}(T)=s_{0}\frac{\langle\bar{q}q\rangle}{\langle0|\bar{q}q|0\rangle}\Big{(}1-\frac{(m_{c}+m_{s})^{2}}{s_{0}}\Big{)}+(m_{c}+m_{s})^{2},  \\
\end{equation}
where in the right hand side $s_{0}$ is hadronic threshold at zero
temperature: $s_{0}=s(T=0)$.

\section{Numerical analysis of mass and leptonic decay constant}
In this section we present our results for the temperature
dependence of $D_{s0}(2317)$ meson mass and leptonic decay constant.
Performing Borel transformation with respect to $Q_{0}^{2}$ from
both sides of equation (15) and taking the derivative with respect
to $1/M^{2}$ from both sides of obtained expression, and making some
transformations we have
\begin{equation}\label{eqn17}
m^{2}(T)=B(T)/A(T),\\
\end{equation}
\begin{equation}\label{eqn18}
f^{2}(T)=\frac{A(T)}{m^{4}(T)}\exp\Big{(}\frac{m^{2}(T)}{M^{2}}\Big{)},\\
\end{equation}
where
\begin{eqnarray}\label{eqn19}
&&A(T)=\int^{s_{0}(T)}_{(m_{c}+m_{s})^{2}}ds
\Big{(}\rho_{a,pert}(s)+\rho_{1}(s)\Big{)}\exp\Big{(}-\frac{s}{M^{2}}\Big{)}
\nonumber\\&&+\int^{(m_{c}-m_{s})^{2}}_{0}ds
\rho_{s,pert}(s)\exp\Big{(}-\frac{s}{M^{2}}\Big{)}+\Pi_{np}(M^{2},T)
,
\end{eqnarray}
\begin{eqnarray}\label{eqn20}
&&\Pi_{np}(M^{2},T)=m_{c}^{3}\langle\bar{s}s\rangle
e^{-\beta}\Big{[}1-\frac{3}{2}\varepsilon+\frac{1}{2}\beta\varepsilon-\beta\varepsilon^{2}\Big{(}1-\frac{1}{2}\beta\Big{)}+\frac{1}{2}\varepsilon^{3}\Big{(}1+\beta-2\beta^{2}+\frac{1}{3}\beta^{3}\Big{)}\Big{]}
\nonumber\\&&+\frac{1}{12}\Big{\langle}\frac{\alpha_{s}G^{2}}{\pi}\Big{\rangle}
m_{c}^{2}e^{-\beta}\Big{[}1-3\varepsilon\Big{(}1-\frac{8}{3}\beta+\beta^{2}-2\beta(\ln(\beta\varepsilon)+\gamma-1)+\beta^{2}\Big{(}\ln(\beta\varepsilon)+\gamma-\frac{3}{2}\Big{)}\Big{)}\Big{]}
\nonumber\\&&+\frac{1}{2}M_{0}^{2}m_{c}\beta\langle\bar{s}s\rangle
e^{-\beta}\Big{[}1-\frac{1}{2}\beta-2\varepsilon\Big{(}1-\frac{3}{4}\beta\Big{(}1-\frac{1}{9}\beta\Big{)}\Big{)}\Big{]}
\nonumber\\&&-\frac{4}{81}\pi\rho\alpha_{s}\langle\bar{s}s\rangle^{2}\beta
e^{-\beta}(12-3\beta-\beta^{2}),
\end{eqnarray}
where $\beta=m_{c}^{2}/M^{2}$ and
$B(T)=-m_{c}^{2}\frac{dA(T)}{d\beta}$.

\begin{table}[h]
\begin{center}
\caption{QCD input parameters used in the analysis.}
\begin{tabular}{ll}
\\
\\
\hline
  \\
Parameters& References\\
\\
\hline
\\
$m=2317$ MeV&\cite{30}\\
$m_{s}=120$ MeV &\cite{30}\\
$m_{c}=1.47$ GeV&\cite{12,30}\\
$f=201$ MeV&\cite{12,30}\\
$\rho=4$&\cite{23,26}\\
$\langle0|\overline{q}q|0\rangle=-0.014~$GeV$^3$&\cite{18}\\
$\langle0|\frac {1}{\pi}\alpha_{s}G^{2}|0\rangle=0.012~$GeV$^4$&\cite{18}\\
$\alpha_s\langle0|\overline{q}q|0\rangle^{2}=5.8\times10^{-4}~$GeV$^6$&\cite{12}\\
$M_{0}^{2}=0.8~$GeV$^2$&\cite{12}\\
$\langle0|\overline{s}s|0\rangle=0.8\langle0|\overline{q}q|0\rangle$&\cite{12}\\
\\
\hline
\end{tabular}
\end{center}
\end{table}
For the numerical evolution of the above sum rule, we use QCD impute
parameters showed in Table 1. The criterion we adopt here is to fix
in such a way so as to reproduce the zero temperature values of
meson mass and leptonic decay constant. $D_{s0}(2317)$ meson mass as
a function of temperature are shown in Fig.1, Fig.3 and Fig.5 at
continuum threshold values $s_{0}=7.5; 8.0; 8.5~GeV^{2}$
respectively. As seen, mass decreases with increasing temperature
and mesons lose approximately $10-15$ percent of its mass at
$T=150~MeV$ temperature. The results for leptonic decay constants
are shown in Fig.2, Fig.4 and Fig.6 at continuum threshold values
$s_{0}=7.5; 8.0; 8.5~GeV^{2}$ respectively. As can be seen $f$
decreases with increasing temperature and vanishes approximately at
critical temperature.  This situation may be interpreted as a signal
for deconfinement and agrees with light and heavy mesons
investigations \cite{16}, \cite{23}. Numerical analysis shows that
the temperature dependence of $f$ is the same, when $M^{2}$ changes
between $1.5~GeV^{2}$ and $3~GeV^{2}$ at fixed values of continuum
threshold. Obtained results can be used for interpretation heavy ion
collision experiments.
\section{Acknowledgement}
The authors much pleasure to thank T. M. Aliev and A. \"{O}zpineci
for useful discussions. This work is supported by the Scientific and
Technological Research Council of Turkey (TUBITAK), research project
no.105T131.
\begin{figure}[h]
\centerline{\epsfig{figure=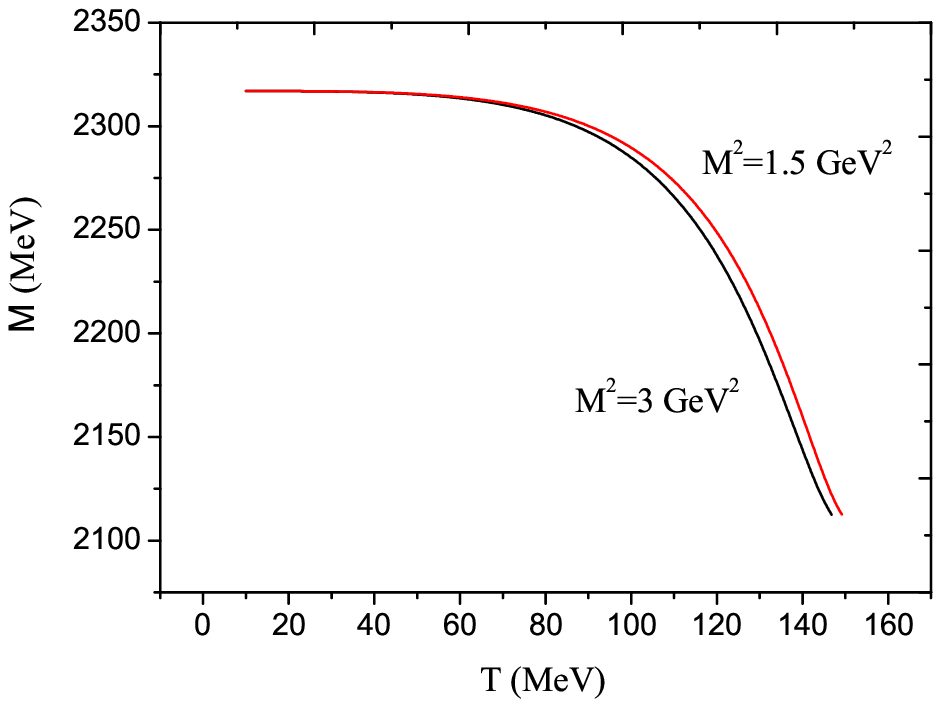,height=80mm}}\caption{Temperature
dependence of meson mass at $s_{0}=7.5~GeV^2$.} \label{BRmh0}
\end{figure}
\begin{figure}[h]
\centerline{\epsfig{figure=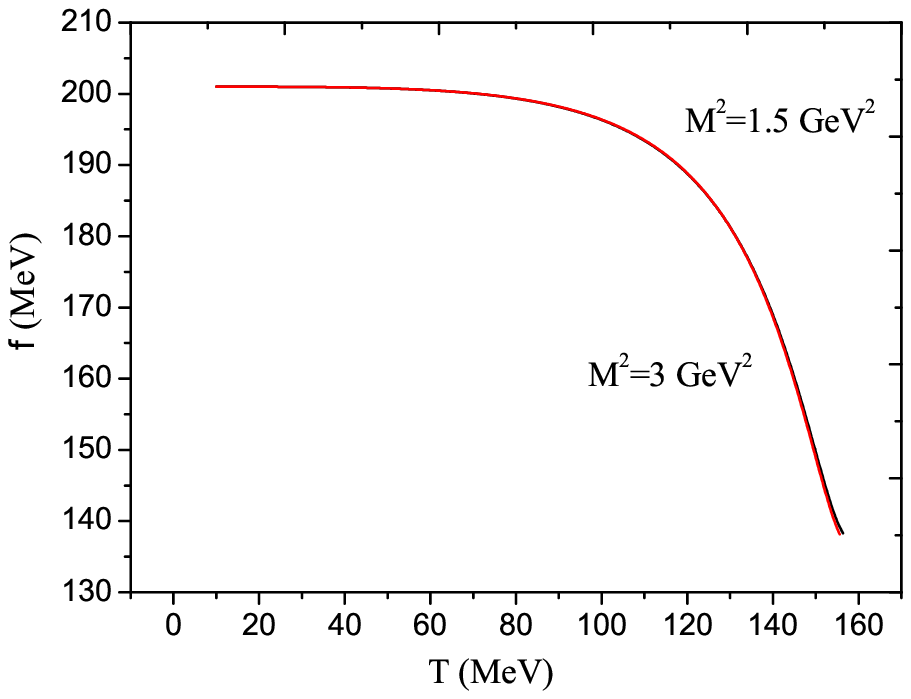,height=80mm}}\caption{
Temperature dependence of leptonic decay constants at
$s_{0}=7.5~GeV^2$.} \label{BRmh0}
\end{figure}
\begin{figure}[h]
\centerline{\epsfig{figure=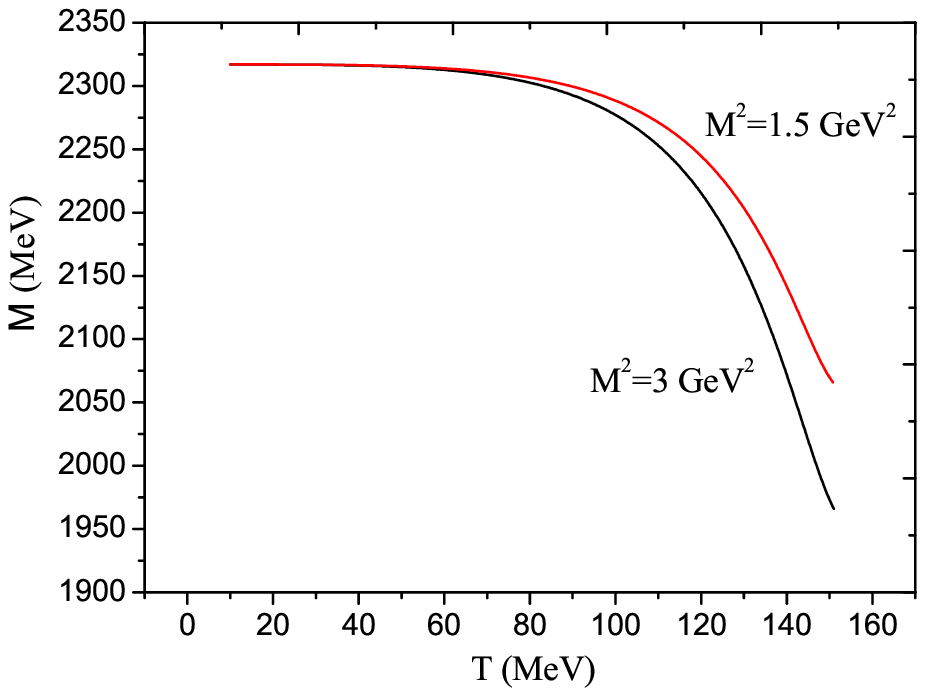,height=80mm}}\caption{
Temperature dependence of meson mass at $s_{0}=8.0~GeV^2$.}
\label{BRmh0}
\end{figure}
\begin{figure}[h]
\centerline{\epsfig{figure=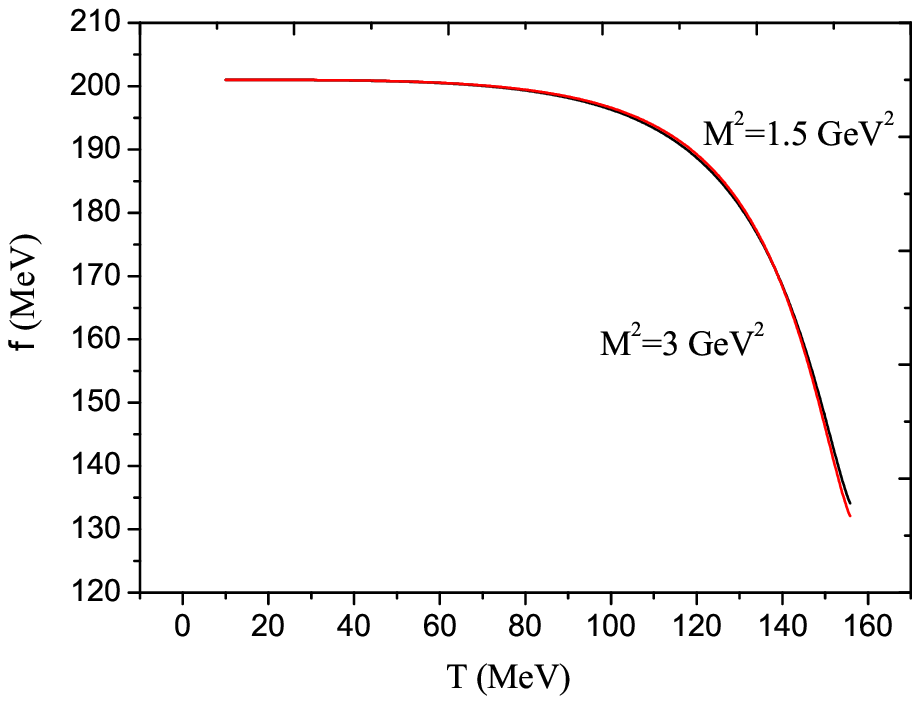,height=80mm}}\caption{
Temperature dependence of leptonic decay constants at
$s_{0}=8.0~GeV^2$.} \label{BRmh0}
\end{figure}
\begin{figure}[h]
\centerline{\epsfig{figure=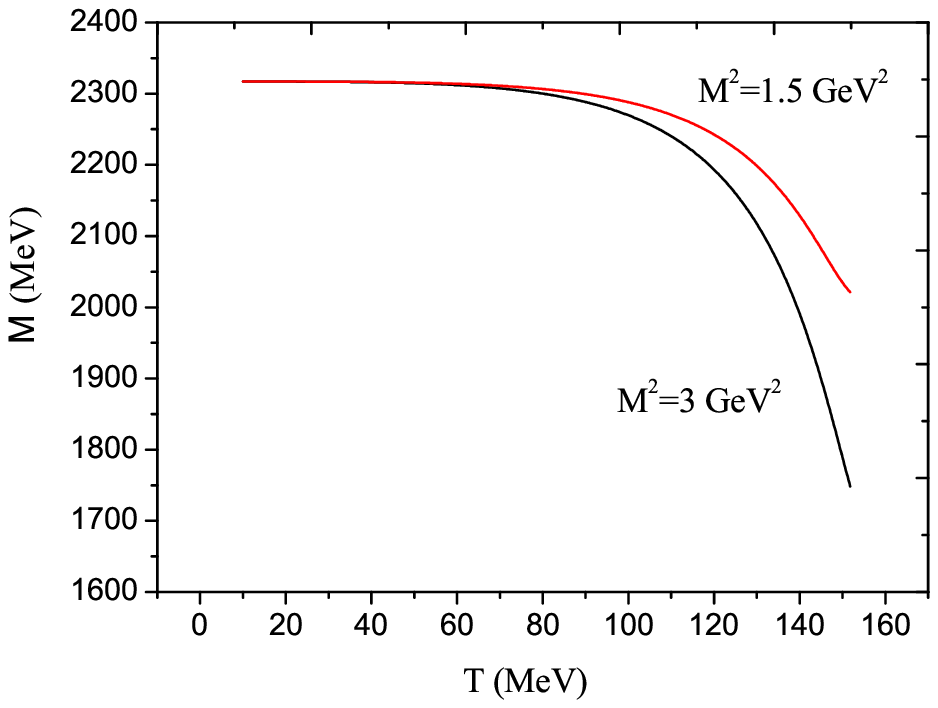,height=80mm}}\caption{
Temperature dependence of meson mass at $s_{0}=8.5~GeV^2$.}
\label{BRmh0}
\end{figure}
\begin{figure}[h]
\centerline{\epsfig{figure=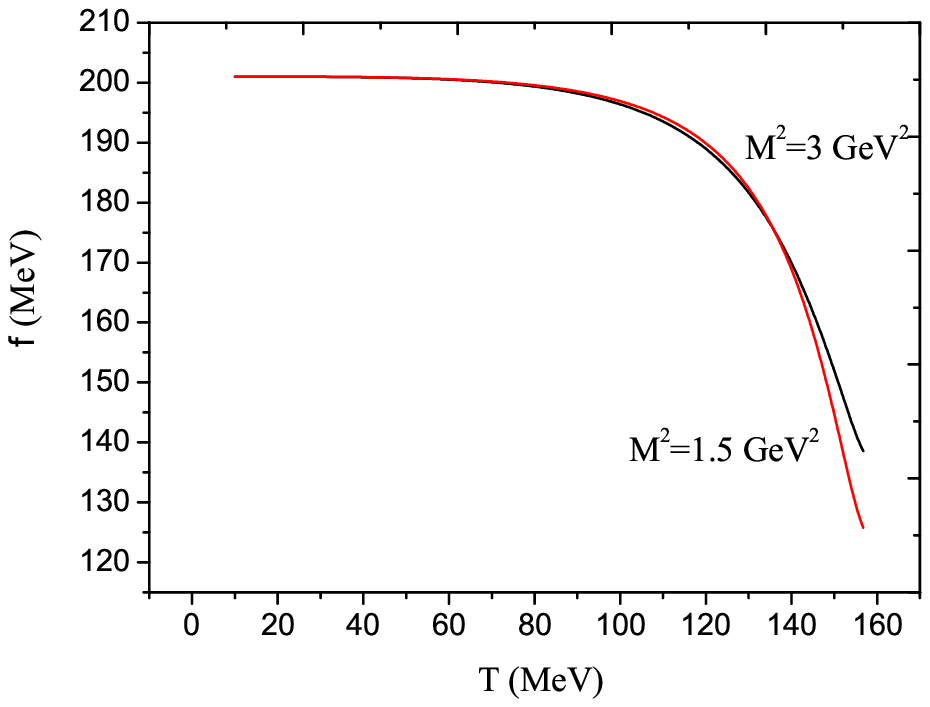,height=80mm}}\caption{
Temperature dependence of leptonic decay constants at
$s_{0}=8.5~GeV^2$.} \label{BRmh0}
\end{figure}

\begin{thebibliography}{1}
%
\bibitem{1}BaBar Collaboration, B. Aubert et al., Phys. Rev. Lett. 90, 242001 (2003).
%
\bibitem{2}CLEO collaboration, D. Besson et al., Phys. Rev. D68, 032002 (2003).
%
\bibitem{3}BELLE Collaboration, P. Krokovny et al., Phys. Rev. Lett. 91, 262002 (2003).
%
\bibitem{4}S. Godfrey and N. Isgur, Phys. Rev. D32, 189 (1985); S. Godfrey and R. Kokoski, Phys. Rev. D43, 1679 (1991); M. Di Pierro and E. Eichten, Phys. Rev. D64, 114004 (2001).
%
\bibitem{5}P. Colangelo and F. De Fazio, Phys. Lett. B 570, 180 (2003).
P. Colangelo, F. De Fazio, A Ozpineci Phys.Rev.D72:074004,2005
%
\bibitem{6}R. N. Cahn and J. D. Jackson, Phys. Rev. D68, 037502 (2003).
%
\bibitem{7}S. Godfrey, Phys. Lett. B 568, 254 (2003).
%
\bibitem{8}A. Datta and P. J. O'Donnell, Phys. Lett. B 572, 164 (2003).
%
\bibitem{9}C. H. Chen and H. n. Li, Phys. Rev. D 69, 054002 (2004).
%
\bibitem{10}M. Q. Huang,  Phys.Rev. D69 (2004) 114015.
%
\bibitem{11}Y. B. Dai, C. S. Huang, C. Liu, and S. L. Zhu, Phys. Rev. D 68, 114011 (2003).
%
\bibitem{12}S. Narison, Phys. Lett. B 520, 115 (2001); S. Narison, Phys. Lett. B 605, 319 (2005).
%
\bibitem{13}T. Barnes, F.E. Close, and H.J. Lipkin, Phys. Rev. D68, 054006 (2003).
%
\bibitem{14}A.P. Szczepaniak, Phys. Lett. B567, 23 (2003).
%
\bibitem{15}H.Y. Cheng and W.S. Hou, Phys. Lett. B566, 193 (2003); K. Terasaki, Phys. Rev. D68,
011501 (2003).
%
\bibitem{16}T. Browder, S. Pakvasa, and A.A. Petrov, Phys. Lett. B 578, 363 (2004).
%
\bibitem{17}W. A. Bardeen, E. J. Eichten, and C. T. Hill, Phys. Rev. D68, 054024 (2003).
%
\bibitem{18}M.A. Shifman, A.I. Vainstein and V.I. Zakharov, Nucl. Phys. B147, 385 (1979), M.A. Shifman, A.I. Vainstein and V.I. Zakharov, Nucl. Phys. B147, 448 (1979).
%
\bibitem{19}A. I. Bochkarev and M. E. Shaposhnikov, Nucl. Phys. B268,  220 (1986).
%
\bibitem{20}E.V. Shuryak, Rev. Mod. Phys., 65, 1, (1993).
%
\bibitem{21}T. Hatsuda, Y. Koike, S. H. Lee, Nucl. Phys. B394, 221 (1993).
%
\bibitem{22}S. Mallik  Phys. Lett. B416, 373 (1998); S. Mallik and K. Mukherjee, Phys. Rev. D58, 096011
(1998).
%
\bibitem{23}C. A., Dominguez, M. Loewe,  J.C. Rojas, JHEP 08, 040, (2008);  E. V. Veliev, G. Kaya Eur. Phys. J. C 63, 87 (2009);  C. A., Dominguez, M. Loewe,  J.C. Rojas, Y. Zhang, arXiv:0908.2709v2.
%
\bibitem{24}E.V. Veliev, J. Phys. G: Nucl. Part. Phys., 35, 035004(2008); E. V. Veliev, T. Aliev, J. Phys. G: Nucl. Part. Phys. 35, 125002, (2008).
%
\bibitem{25}A. Das,  Finite Temperature Field Theory, World Scientific(1999).
%
\bibitem{26}C. A., Dominguez and N., Paver, Phys. Lett. B318, 629 (1993).
%
\bibitem{27}J. Gasser and H. Leutwyler, Phys. Lett. B184, 83 (1987).
%
\bibitem{28}P. Gerber and H. Leutwyler, Nucl. Phys. B231, 387 (1989).
%
\bibitem{29}D.E. Miller, Acta Phys. Pol. B 28, 2937 (1997), D.E. Miller, arXiv: hep-ph/0008031.
%
\bibitem{30}PDG 2008, C. Amsler, et al.., Phys. Lett B 667, 1 (2008).
%
\end{thebibliography}
\end{document}